\documentclass[prl,aps,epsf,showpacs,twocolumn]{revtex4-1}
\usepackage{times}
\usepackage{graphicx}
\usepackage{float}
\usepackage{latexsym,amsmath,amssymb,bm,euscript}
\usepackage{color}
\usepackage{subfigure}
\usepackage{epstopdf}
\usepackage[colorlinks=true,linkcolor=blue,citecolor=blue,urlcolor=blue]{hyperref}
\usepackage{hyperref}
\usepackage{ulem}
\usepackage[dvipsnames]{xcolor}
\usepackage{dcolumn}     
\usepackage{slashbox}     

\begin{document}

\title{Nematic and Antiferromagnetic Quantum Criticality in a Multi-Orbital Hubbard Model for Iron Pnictides}

\author{Wen-Jun Hu$^{1}$}
\author{Haoyu Hu$^{1}$} 
\author{Rong Yu$^{2}$}
\author{Hsin-Hua Lai$^{1}$}
\author{Luca F. Tocchio$^{3}$}
\author{Federico Becca$^{4}$}
\email{fbecca@units.it}
\author{Qimiao Si$^{1}$}
\email{qmsi@rice.edu}
\affiliation{
$^1$ Department of Physics and Astronomy \& Rice Center for Quantum Materials, Rice University, Houston, Texas 77005, USA \\
$^2$ Department of Physics and Beijing Key Laboratory of Opto-electronic Functional Materials and Micro-nano Devices, Renmin University of China, Beijing 100872, China\\
$^3$ Institute for condensed matter physics and complex systems, DISAT, Politecnico di Torino, I-10129, Italy \\
$^4$ Department of Physics, University of Trieste, Strada Costiera 11, Trieste 34151, Italy
}

\begin{abstract}
The extent to which quantum criticality drives the physics of iron pnictides is a central question in the field. Earlier theoretical considerations were based 
on an effective field theory, and the proposed realization in P-doped iron arsenides has received extensive experimental evidence. To connect the quantum 
critical behavior with the underlying electronic physics, it is important to analyze it within microscopic models. Here, we do so for a multi-orbital model 
containing both Hubbard and Hund's interactions, by a variational Monte Carlo method based on Jastrow-Slater wave functions that allow for a non-perturbative 
treatment of electron correlations. We find strong evidence for the existence of a {\it unique} quantum critical point, where both nematic and $(\pi,0)$ 
antiferromagnetic orders develop together, in the bad-metal regime of the phase diagram. Implications of our results for the iron-based superconductivity are 
discussed.
\end{abstract}


\maketitle

{\it Introduction --}
Iron-based superconductivity develops in correlated bad metals near an antiferromagnetic (AFM) order~\cite{Kamihara2008,PCDai_review12,Hir.16,Si2016}.
Furthermore, the critical temperature $T_{c}$ for the onset of superconductivity has a dome-shaped dependence as a function of doping or external pressure.
These features share a considerable degree of similarity with those of the high-$T_{c}$ cuprates~\cite{Lee_rmp.06,dagotto1994,Ima.98} and heavy-fermion 
systems~\cite{Coleman-Nature,Si_Science10}. Like in the latter systems, the role of quantum criticality is an important question in the physics of both 
the normal and superconducting states of the iron-based systems.

From the experimental point of view, the most extensive studies on quantum criticality in the iron-based materials have taken place in the so-called $122$ 
family, with the parent compound BaFe$_{2}$As$_{2}$. This family provides a unique opportunity to study the evolution of the electronic properties over a 
wide range of the phase diagram. In particular, the phosphorus-substituted system, BaFe$_{2}$(As$_{1-x}$P$_{x}$)$_{2}$, is rather clean since P-substitution 
does not induce appreciable scattering~\cite{HShishido2010,Ana.10a,CJvdBeek2010}. Extensive experimental observations (for earlier reviews, see 
Ref.\,\cite{Abr.11,shibauchi2014}) have provided strong evidence of quantum criticality in this materials series~\cite{Jia.09,SKasahara2010,hayes2016,dai2018}
and in related iron-based systems~\cite{Cru.10,Kuo.16}.

From the theoretical perspective, quantum criticality in iron pnictides was proposed within an effective field theory that contains both AFM and nematic order 
parameters~\cite{dai2009,wu2016}. This theoretical analysis provided the basis for the early proposal of realizing quantum criticality in P-doped iron 
arsenides~\cite{dai2009}. A quantum critical point (QCP) corresponds to a special class of second-order phase transitions that takes place at zero temperature,
typically in a material in which the critical temperature has been driven to zero by non-thermal parameters~\cite{sachdev2011quantum,Coleman-Nature,Si_Science10}.
Whether a putative QCP exists in the phase diagram of the iron-pnictide materials or not remains debated. Nevertheless, it is widely believed that a QCP is 
hidden beneath the superconducting dome; deepening our understandings of the physics regarding a QCP may then shed light on the mechanism of unconventional 
superconducting pairing. In order to address the physics of these compounds, {\it ab-initio} calculations based on the standard local-density approximation have 
been performed~\cite{lebegue2007,singh2008,cao2008}. While this approach was useful to highlight the rather complex band-structure, with several $d$ orbitals 
participating in the low-energy properties, it does not properly treat the effects of strong correlations and electron pairing.

To address the physics of quantum phase transitions in microscopic studies requires a systematic treatment of correlated multi-orbital models with both Hubbard 
and Hund's interactions. The bad-metal nature of the normal state implies that the interactions are of the same order as the kinetic energy and, thus, the need 
to treat correlated states beyond perturbative approaches. In particular, Hund's interactions are important for the correlation effects in this regime. 
Additionally, the entwining of AFM and nematic correlations highlights the need to address spatial correlations.

In this letter, we aim at identifying the QCP and studying the physics around it within a multi-orbital model that contains both Hubbard (intra-orbital $U$ and 
inter-orbital $U'$) and Hund's ($J_H$) interactions. We perform variational Monte Carlo (VMC) simulations with Jastrow-Slater states. While VMC analyses have 
previously been carried out in multi-orbital Hubbard models~\cite{fan2011,watanabe2012,Tocchio2016,Franco2018}, in our study we incorporate a spin Jastrow factor,
in order to improve the description of correlation effects that originate from the Hund's term. At large electron-electron interactions, we find a stable AFM order with 
pitch vector ${\bf Q}=(0,\pi)$ [or equivalently $(\pi,0)$] coexisting with a nematic order, i.e., a spontaneous breaking of $\pi/2$ rotations of the lattice. 
As the interaction decreases, a {\it unique} phase transition takes place, where both antiferromagnetism and nematicity disappear. This represents the most 
important result of the present work. Furthermore, by analysing the behavior of the double occupation, we identify the location of the metal-insulator (Mott) 
transition, which appears not too far above the AFM-nematic QCP.

\begin{figure}[t]
\includegraphics[width=\columnwidth]{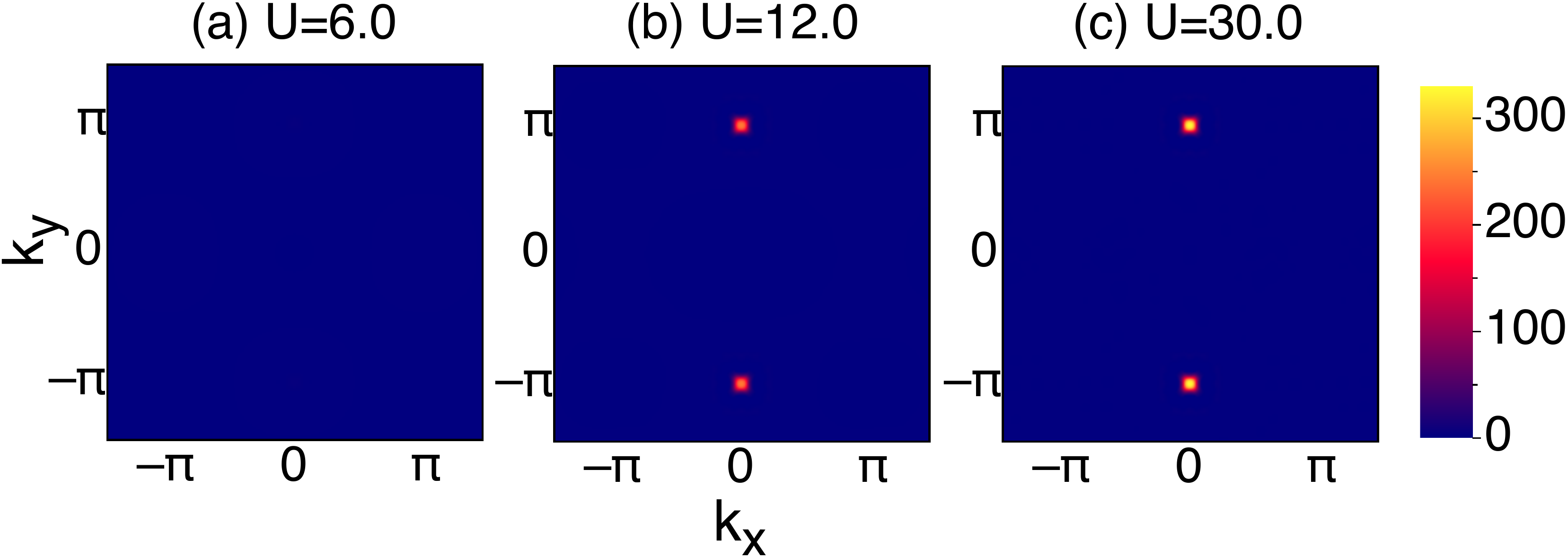}
\includegraphics[width=\columnwidth]{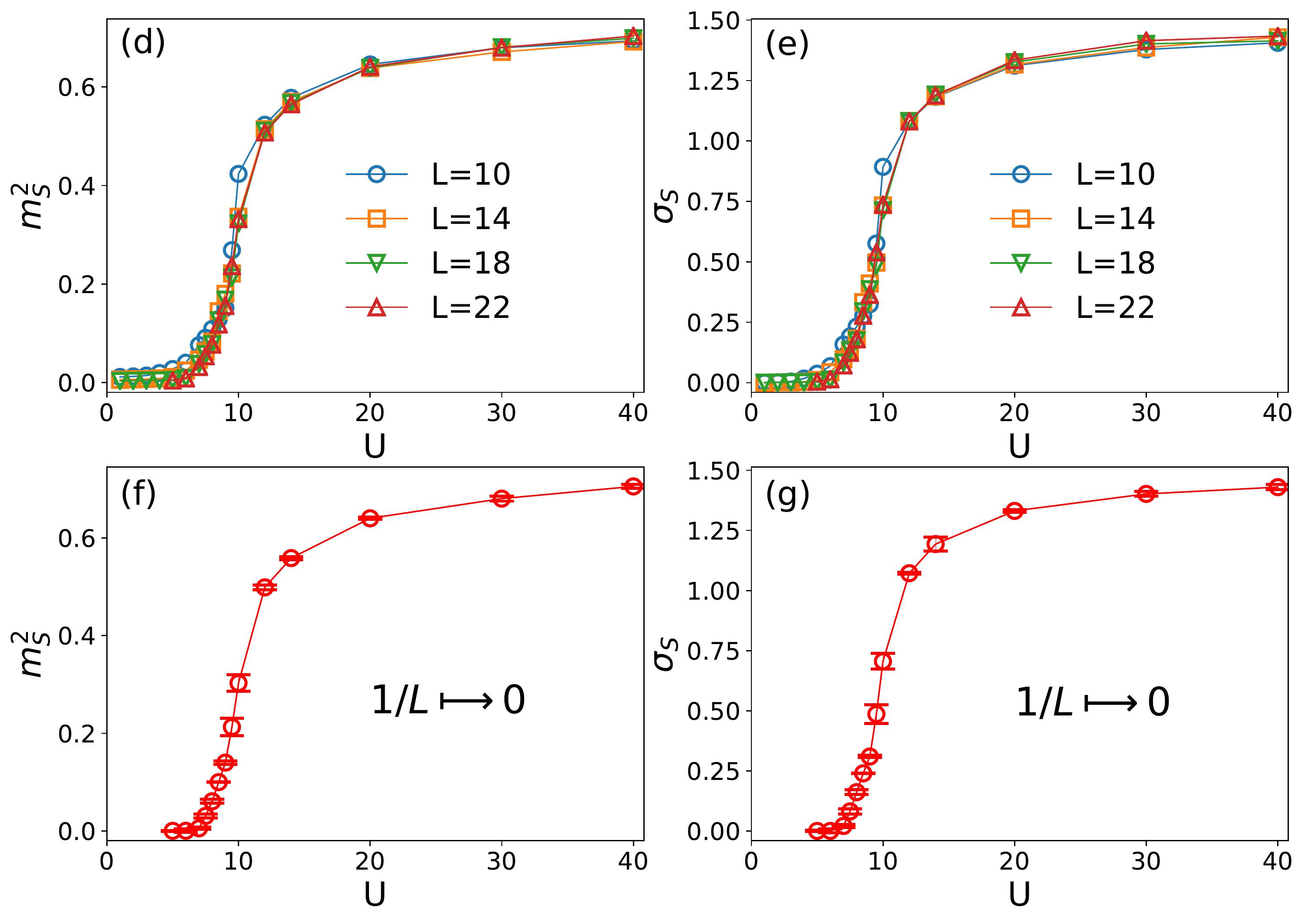}
\caption{Upper panels: Spin structure factor $S(q)$ at $U=6.0$ (a), $12.0$ (b), and $30.0$ (c) on a $L=22$ lattice, for the 2-orbital Hubbard model at half filling 
with the Hund's coupling $J_H=0.1U$. Middle panels: Magnetic ($m^2_S$) and nematic ($\sigma_S$) order parameters as a function of $U$ for $L=10$, $14$, $18$, and 
$22$ clusters. Lower panels: Thermodynamic extrapolations ($L \mapsto \infty$) for $m^2_S$ and $\sigma_S$.}
\label{sno2}
\end{figure}

\begin{figure}[t]
\includegraphics[width=\columnwidth]{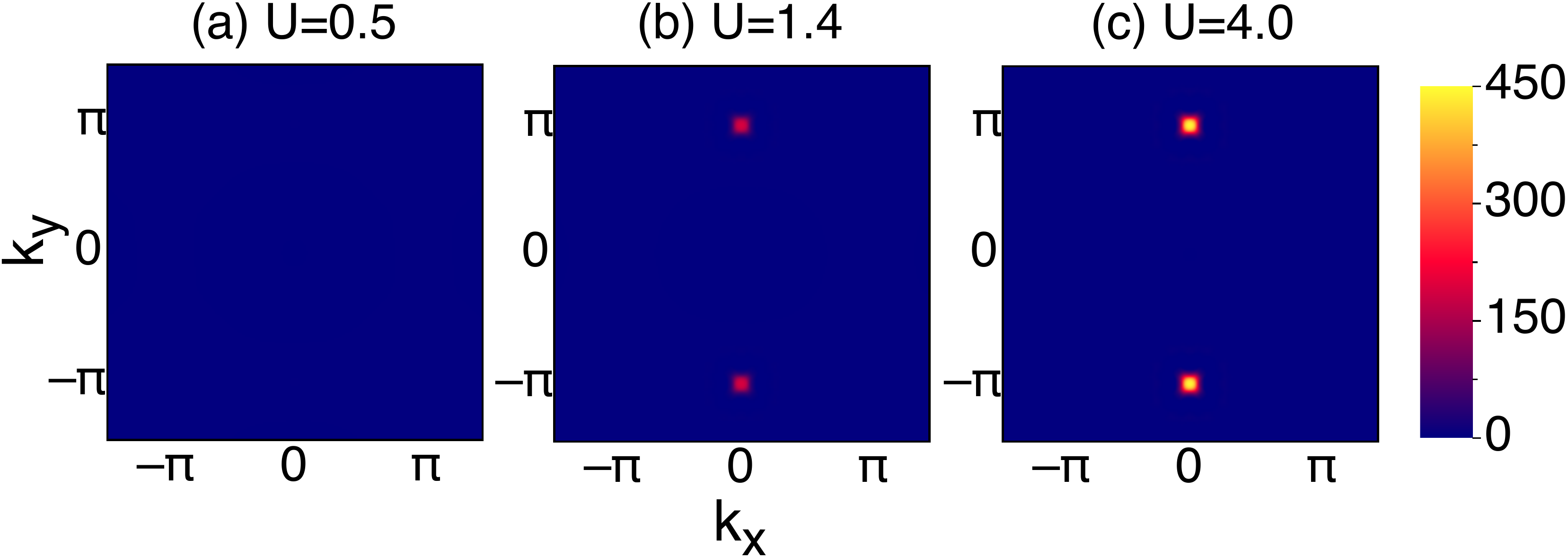}
\includegraphics[width=\columnwidth]{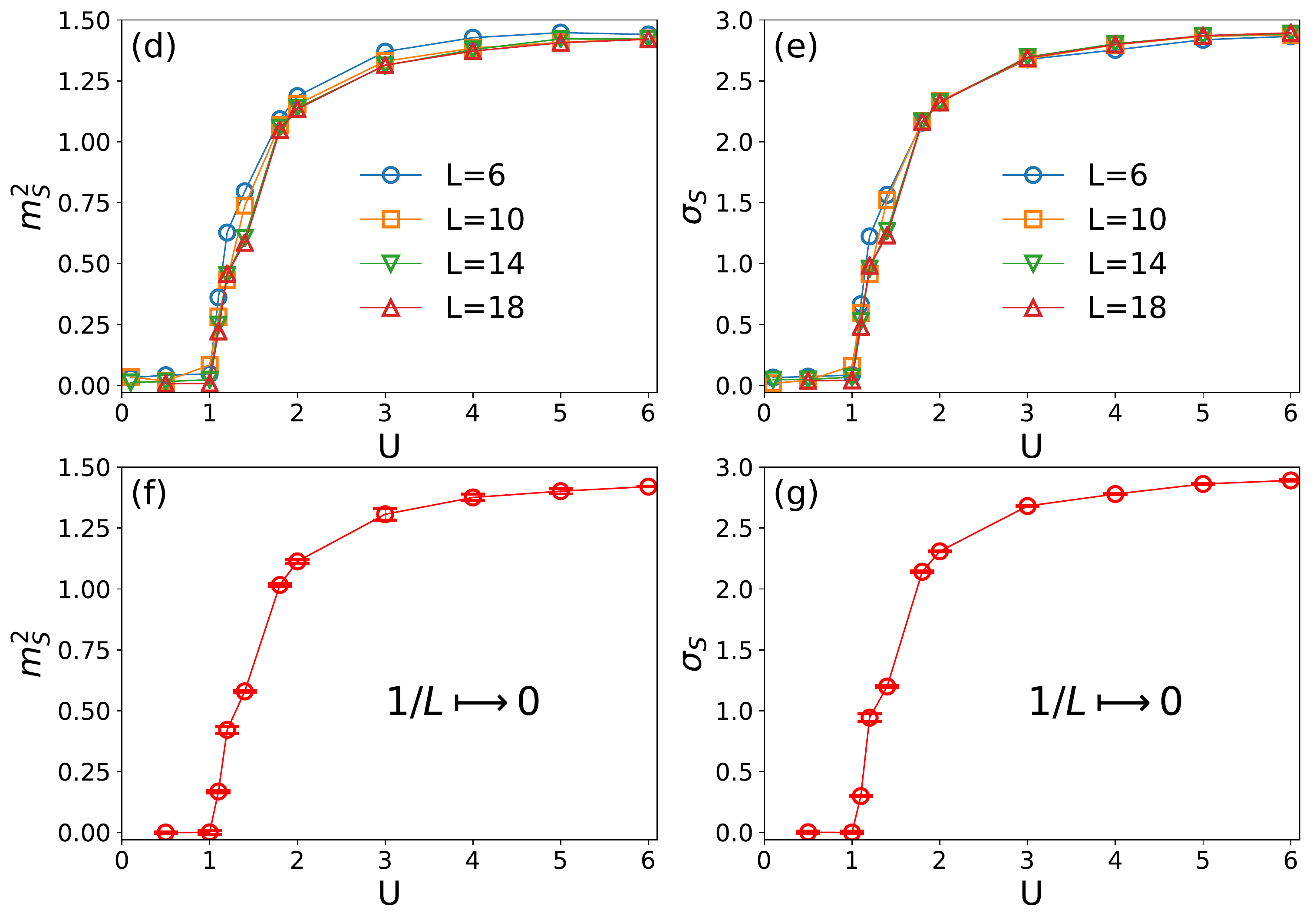}
\caption{Upper panels: Spin structure factor $S(q)$ at $U=0.5$ (a), $1.4$ (b), and $4.0$ (c) on a $L=18$ lattice for the 3-orbital Hubbard model at half filling 
with the Hund's coupling $J_H=0.1U$. Middle panels: Magnetic ($m^2_S$) and nematic ($\sigma_S$) order parameters as a function of $U$ for $L=6$, $10$, $14$, and 
$18$ clusters. Lower panels: Thermodynamic extrapolations ($L \mapsto \infty$) for $m^2_S$ and $\sigma_S$.}
\label{sno3}
\end{figure}

{\it Models and Methods --}
Theoretically, in the iron pnictides, there are six electrons occupying the $3d$ Fe orbitals. In these materials, it is fundamental to include this multiplicity 
of the involved orbitals~\cite{Hir.16,Si2016}. The most generic Hamiltonian that describes $N_{\rm orb}$ orbitals with inter- and intra-orbital interactions is 
given by:
\begin{eqnarray}
 \label{mohm} 
{\cal H} &=& -\sum_{i,j,\alpha,\beta,\sigma} t^{\alpha,\beta}_{i,j} c^\dag_{\alpha,i,\sigma} c^{\phantom{\dag}}_{\beta,j,\sigma}
+ \sum_{i,\alpha} \Delta^{\rm cf}_{\alpha} n_{\alpha,i} \nonumber \\
&+& U \sum_{i,\alpha} n_{\alpha,i,\uparrow}n_{\alpha,i,\downarrow} + 
\left ( U^\prime-\frac{J_{\rm H}}{2} \right ) \sum_{i,\alpha<\beta} n_{\alpha,i} n_{\beta,i} \nonumber \\
&-& 2J_{\rm{H}} \sum_{i,\alpha<\beta} {\bf S}_{\alpha,i}\cdot {\bf S}_{\beta,i} \nonumber \\
&+& J_{\rm{H}} \sum_{i,\alpha<\beta} 
c^\dagger_{\alpha,i,\uparrow}c^\dagger_{\alpha,i,\downarrow}c^{\phantom{\dag}}_{\beta,i,\downarrow}c^{\phantom{\dag}}_{\beta,i,\uparrow}+ \textrm{h.c.},
\end{eqnarray}
where $c^{\phantom{\dag}}_{\alpha,i,\sigma}$ is the electronic annihilation operator with spin $\sigma$ on orbital $\alpha$ and site $i$, and the density operator 
is $n_{\alpha,i,\sigma}= c^\dag_{\alpha,i,\sigma} c^{\phantom{\dag}}_{\alpha,i,\sigma}$, with $n_{\alpha,i} = \sum_{\sigma} n_{\alpha,i,\sigma}$; moreover,
${\bf S}_{\alpha,i}=\frac{1}{2}\sum_{\sigma,\sigma'}c^{\dag}_{\alpha,i,\sigma}\tau_{\sigma,\sigma'}c^{\phantom{\dag}}_{\alpha,i,\sigma'}$, $\tau$ being the Pauli 
matrices, are the spin operators. Here, $t^{\alpha,\beta}_{i,j}$ are the intra- ($\alpha=\beta$) and inter-orbital ($\alpha \neq \beta$) hopping amplitudes, 
respectively; $\Delta^{\rm cf}_{\alpha}$ denotes the crystal field; $U$ and $U'=U-2J_H$ are the intra- and inter-orbital on-site Hubbard interactions, while $J_H$ 
is the Hund's coupling~\cite{georges2013}. Hopping amplitudes and crystal fields can be estimated by density functional theory through a comparison with the 
experimental data of the Fermi surface. While a full description of the $3d$ shell of Fe orbitals would imply $N_{\rm orb}=5$~\cite{Hir.16,Si2016}, there are 
simplified versions restricting to $N_{\rm orb}=3$ or $2$ orbitals. While the former one includes all the $t_{2g}$ orbitals (with $xz$, $yz$, and $xy$ 
symmetry)~\cite{dagotto2010}, the latter one restricts to the case in which only $xz$ and $yz$ orbitals are kept in the low-energy description of iron 
pnictides~\cite{zhang2008,dagotto2009}. We mention that the 3-orbital model contains two Fe atoms per unit cell, but full translational invariance can be 
recovered after considering $c^\dag_{\alpha,i,\sigma} \to (-1)^{{\rm R}_i} c^\dag_{\alpha,i,\sigma}$ for $\alpha=xz$ and $yz$ (in the 2-orbital case, this
transformation does not spoil translational invariance). For the 2-orbital model, there are no crystal fields and the actual values of the hopping amplitudes 
are $t^{1,1}_{i,i+x}=t^{2,2}_{i,i+y}=-1.3$, $t^{1,1}_{i,i+y}=t^{2,2}_{i,i+x}=1.0$, $t^{\alpha,\alpha}_{i,i+x+y}=t^{\alpha,\alpha}_{i,i+x-y}=-0.85$, 
and $t^{1,2}_{i,i+x+y}=-t^{1,2}_{i,i+x-y}=0.85$ (with symmetric amplitudes when exchanging orbital indices)~\cite{zhang2008,dagotto2009}. Instead, for the 
3-orbital model, we take the parameters given in Table I of Ref.~\cite{dagotto2010}.

Within the VMC approach, different trial wave functions can be considered, in order to optimize the expectation value of the total energy. In this letter, we 
will choose the Jastrow-Slater wave functions, defined by~\cite{Capello2005}:
\begin{equation}\label{eq:wf}
|\Psi_{v}\rangle = \mathcal{J}_{s}\mathcal{J}_{n}|\Phi_0\rangle;
\end{equation}
here, the uncorrelated state $|\Phi_0\rangle$ is specified by the following auxiliary (quadratic) Hamiltonian~\cite{Tocchio2016,Franco2018}:
\begin{eqnarray}
{\cal H}_{\rm aux} &=& -\sum_{i,j,\alpha,\beta,\sigma} {\tilde t}^{\alpha,\beta}_{i,j} c^\dag_{\alpha,i,\sigma} c^{\phantom{\dag}}_{\beta,j,\sigma}+
\sum_{i,\alpha,\sigma}\mu_{\alpha}c^\dag_{\alpha,i,\sigma} c^{\phantom{\dag}}_{\alpha,i,\sigma} \nonumber\\
&+&\sum_{i,\alpha}e^{i{\bf Q}\cdot{\bf R}_{i}}\Delta_{\alpha}^{\rm AFM}(c^\dag_{\alpha,i,\uparrow} c^{\phantom{\dag}}_{\alpha,i,\downarrow}
+c^\dag_{\alpha,i,\downarrow} c^{\phantom{\dag}}_{\alpha,i,\uparrow}),
\label{eq:Haux} 
\end{eqnarray}
where ${\tilde t}^{\alpha,\beta}_{i,j}$, $\mu^{\alpha}$, and $\Delta^{\rm AFM}_{\alpha}$ are variational parameters. In particular, the presence of 
$\Delta^{\rm AFM}_{\alpha} \ne 0$ implies magnetic order on each orbital. By choosing ${\bf Q}=(\pi,\pi)$ or $(0,\pi)$ [or equivalently $(\pi,0)$], we can 
have either the N\'eel order or the collinear AFM order (CAFM). To allow for the treatment of the correlation effect associated with the Hund's coupling, 
we go beyond previous studies of multi-orbital Hubbard models~\cite{fan2011,watanabe2012,Tocchio2016,Franco2018} by incorporating the spin Jastrow factor 
in addition to the usual density one:
\begin{eqnarray}
\mathcal{J}_{n} &=& \exp \left [ \frac{1}{2} \sum_{\alpha,\beta,i,j} u^{\alpha,\beta}_{i,j} n_{\alpha,i} n_{\beta,j} \right ], \\
\mathcal{J}_{s} &=& \exp \left [ \frac{1}{2} \sum_{\alpha,\beta,i,j} v^{\alpha,\beta}_{i,j} S^z_{\alpha,i} S^z_{\beta,j} \right ],
\end{eqnarray}
where $u^{\alpha,\beta}_{i,j}$ and $v^{\alpha,\beta}_{i,j}$ are translationally invariant variational parameters, which introduce quantum fluctuations into the 
wave function. In the context of the one-band Hubbard model, how the intersite spin Jastrow factor better captures the spin-spin correlations has been discussed 
before~\cite{Becca.00}. In Eq.~(\ref{eq:Haux}), we have incorporated the magnetic order parameter through $\Delta_{\alpha}^{\rm AFM}$, which couples to the 
spins perpendicular to the $z$ direction. This improves the description of the ordered magnetic state, given that the spin fluctuations produced by the spin 
Jastrow factor are orthogonal to the direction of the ordered moment in the uncorrelated wavefunction~\cite{Hu.15}.

We use the stochastic reconfiguration optimization method~\cite{sorella} to optimize the variational parameters in both the auxiliary Hamiltonian and the Jastrow 
factors, in order to find the energetically favored state in the VMC scheme. In the following, we allow for states that break rotational symmetry in spin and 
lattice, but always keep the translational symmetry. We emphasize that the optimization includes the effective hopping amplitudes ${\tilde t}^{\alpha,\beta}_{i,j}$, 
which is also allowed to break the C$_4$ rotational symmetry and this is crucial to realize the nematic order. In addition, the Jastrow factor is optimized up to 
the next-nearest neighbor in order to limit the number of variational parameters. In the following, we denote with $L$ the linear dimension of the system, 
corresponding to a $N=L\times L$ total number of sites.

\begin{figure}[t]
\includegraphics[width=\columnwidth]{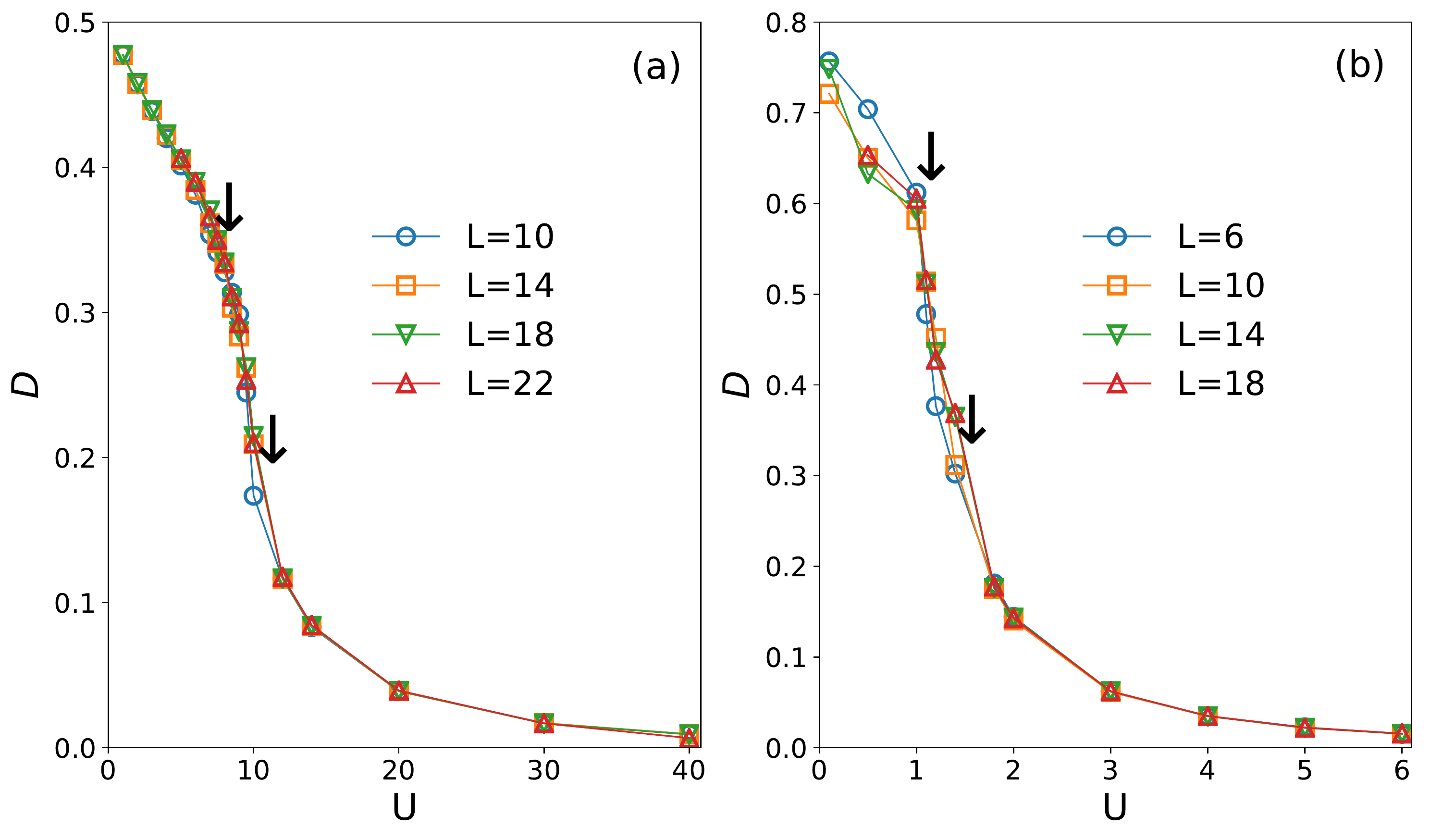}
\caption{(a) The doublon density as a function of $U$ on $L=10$, $14$, $18$, and $22$ clusters for the 2-orbital Hubbard model at half filling with the Hund's 
coupling $J_H=0.1U$. (b) The doublon density as a function of $U$ on $L=6$, $10$, $14$, and $18$ clusters for the 3-orbital Hubbard model at half filling with 
the Hund's coupling $J_H=0.1U$. Black vertical arrows denote kinks related to the AFM-nematic transition and the metal-insulator transition, where the doublon 
density shows a bending.}
\label{doublon}
\end{figure}

{\it Results --}
We have performed VMC simulations for both the 2- and the 3-orbital Hubbard model at half filling~\cite{notefilling} with the Hund's coupling $J_H=0.1U$, and 
considered both the N\'eel ordered phase and the CAFM one in the auxiliary Hamiltonian of Eq.~(\ref{eq:Haux}). After optimization, we find that the CAFM order 
has lower energy than the N\'eel one, especially in the large $U$ region, for both the 2- and the 3-orbital Hubbard model. Therefore, in the following, we show 
numerical results only for the variational wave function with the CAFM order.

In order to identify the magnetic order, we compute the spin structure factor as
\begin{equation}\label{m2}
S(q)=\frac{1}{N}\sum_{i,j} \langle{\bf S}_{i}\cdot {\bf S}_{j}\rangle e^{i{\bf q}\cdot({\bf R}_i-{\bf R}_j)},
\end{equation}
where $\langle{\bf S}_{i}\cdot {\bf S}_{j}\rangle$ is the spin-spin correlation function. In addition, we introduce the nematic order parameter $\sigma_S$ to 
assess the rotational symmetry breaking (i.e., the bond nematicity):
\begin{equation}\label{sigma1}
\sigma_S=\frac{1}{N}\sum_{i}[\langle {\bf S}_{i} \cdot {\bf S}_{i+x}\rangle - \langle {\bf S}_{i} \cdot {\bf S}_{i+y}\rangle],
\end{equation}
The results are reported in Figs.~\ref{sno2} and~\ref{sno3} for the 2- and the 3-orbital Hubbard model, respectively. For large electron-electron interactions,
a sharp peak in the spin structure factor is clearly visible at ${\bf Q}=(0,\pi)$, typical of CAFM order. In order to assess the presence of a true long-range
order, we define the magnetic order parameter as $m^2_S=S(0,\pi)/N$, which is also reported in Figs.~\ref{sno2} and~\ref{sno3}, as a function of the interaction 
$U$ on different lattice sizes for both the 2- and the 3-orbital Hubbard model. It is clearly shown that a magnetic phase transition occurs between a CAFM 
ordered phase and a nonmagnetic one in both cases. According to the finite size scaling, we locate the magnetic transition points at $U_c \approx 7.5$ 
for the 2-orbital Hubbard model and at $U_c \approx 1.0$ for the 3-orbital one. Considering the half-bandwidth $W=6$ for the 2-orbital Hubbard 
model~\cite{zhang2008,dagotto2009} and $W=1.5$ for the 3-orbital one~\cite{dagotto2010}, we observe that the magnetic critical points are located at 
$U_c/W \approx 1.25$ for the 2-orbital Hubbard model and $U_c/W \approx 0.666$ for the 3-orbital one. 
 
While a CAFM order with pitch vector ${\bf Q}=(0,\pi)$ would naturally imply nematic order, since correlations along $x$ and $y$ directions are fundamentally 
different, a nematic order may be possible even without CAFM order. In order to clarify this important issue we report the behavior of the nematic order 
parameter $\sigma_S$ as a function of the interaction $U$, for both the 2- and the 3-orbital Hubbard models, see Figs.~\ref{sno2} and~\ref{sno3}. Our numerical 
calculations indicate that the nematic phase transition is concurrent with the magnetic one, strongly suggesting that a pure nematic phase is hardly obtained 
at zero temperature.

\begin{figure}[t]
\includegraphics[width=\columnwidth]{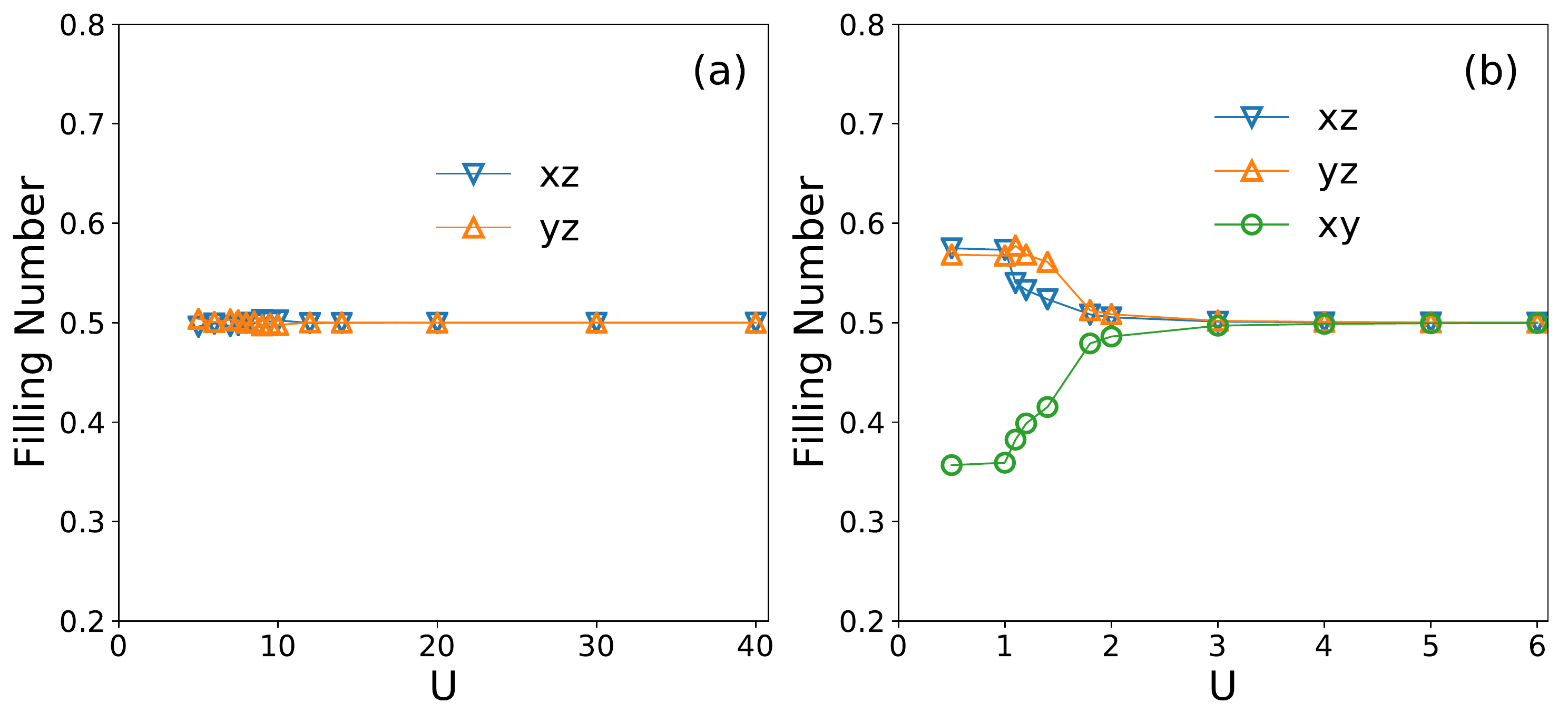}
\caption{(a) The filling numbers of $xz$ and $yz$ orbitals on a $L=22$ lattice for the 2-orbital Hubbard model at half filling with the Hund's coupling $J_H=0.1U$. 
(b) The filling numbers of $xz$, $yz$, and $xy$ orbitals on a $L=18$ lattice for the 3-orbital Hubbard model at half filling with the Hund's coupling $J_H=0.1U$.}
\label{filling}
\end{figure}

Within the multi-orbital models that have been considered so far, the presence of a nonzero AFM order does not necessarily imply the presence of a charge gap.
Therefore, it is important to assess the metallic/insulating behavior of the variational wave function. 
We consider the density of double occupation:
\begin{equation}
D = \frac{1}{N} \sum_{i,\alpha} n_{\alpha,i,\uparrow} n_{\alpha,i,\downarrow}.
\end{equation}
The numerical calculations are shown in Fig.~\ref{doublon} for both the 2- and the 3-orbital Hubbard model. At $U=0$, the doublon density is $0.5$ for the 2-orbital
model and $\approx 0.75$ for the 3-orbital one. When including the Hubbard $U$, $D$ monotonically decreases, showing a small kink connected to the development of 
AFM order and a bending that is directly associated to the opening of the charge gap~\cite{Tocchio2011}, namely to the metal-insulator transition. This can be located 
around $U=11$ and $U=2$ for the 2- and 3-orbital model, respectively.  Since these values are not too much larger than $U_c$ for the AFM-nematic QCP, the system at 
$U_c$ falls in the bad-metal regime. In addition, our analysis highlights the possibility to have an intermediate metallic phase with both AFM and nematic orders, 
sandwiched between a paramagnetic metal at small values of $U$ and an AFM insulator with nematic order at strong coupling. 

Finally, we report the effective filling of each orbital in both cases, see Fig.~\ref{filling}. In the 2-orbital model, both $xz$ and $yz$ are close to half filling 
for every value of the electron-electron interaction; instead, we find that in the 3-orbital model the filling number of the $xy$ orbital is different from the one 
of the $xz$ and $yz$ orbitals, when $U\lesssim 2$, i.e., in the metallic regime.

{\it Discussions --} 
Several points of considerations are in order. First, similar to the Hubbard-Heisenberg model of the one-band case~\cite{Dau.00}, we can introduce $J_1-J_2$ 
interactions into the multi-orbital Hamiltonian Eq.~(\ref{mohm}). We find the results to be qualitatively similar, including the concurrence of the nematic 
and $(\pi,0)$ AFM transitions.

Second, it is instructive to compare the result of our microscopic studies with that of an effective field theory for the AFM and nematic orders. There, the 
coupling between the two degrees of freedom originates from a quartic interaction of the underlying magnetic field. The relevant nature of this coupling turns the 
concurrent transition to be ultimately first order. However, because the coupling is only marginally relevant, it is a weakly first order transition: The jump of 
the order parameter is very small compared to the saturated moment, leaving an extended dynamical range for quantum criticality~\cite{dai2009}. The same conclusion 
is derived by a saddle-point calculation in a large-$N$ limit of the field theory~\cite{wu2016}, and is consistent with experimental 
observations~\cite{shibauchi2014,hayes2016,dai2018,Hu.15a}. In our microscopic study, the numerical results point to a unique QCP for the transition of both orders; 
however, our data do not  rule out the quantum phase transition to be weakly first order, with the order parameters experiencing a small jump.
 
Third, given the purely two-dimensional feature of the model, an infinitesimal temperature completely destroys magnetic order, while nematicity is expected to 
survive up to a critical temperature $T_{\rm n}$, since it is related to a discrete symmetry breaking. In a three-dimensional lattice with nonzero interlayer 
couplings, the AFM order is also stable against thermal fluctuations, and the critical temperature $T_{\rm af}$ will not be linked to $T_{\rm n}$, 
possibly allowing for a pure nematic order, with no long-range antiferromagnetism. 
 
{\it Conclusions --}
We have studied both the 2- and  3-orbital Hubbard models by means of the variational Monte Carlo method. Using optimized Jastrow-Slater wave functions, we 
have calculated both magnetic and nematic order parameters showing the existence of a {\it unique} quantum critical point separating paramagnetic and antiferromagnetic 
(with broken rotational symmetry) phases. Indeed, the size scaling analysis of both magnetic and nematic order parameters indicated the concomitant insurgence of
these orders, limiting the possibility to have a genuine nematic ground state (with no magnetic order) to a tiny region, which is not detected in our calculations.
Furthermore, the behavior of the doublon density suggests that the metal-insulator transition is located slightly inside the antiferromagnetic phase, indicating that 
quantum criticality occurs in the bad-metal regime and that a metallic phase with the CAFM order exists in both the 2- and the 3-orbital Hubbard model.
Our results provide the theoretical basis to understand the quantum criticality observed in the experiments on the iron pnictide 
BaFe$_{2}$As$_2$~\cite{Jia.09,SKasahara2010,hayes2016,dai2018}. In addition, the concurrence of the antiferromagnetic and nematic transitions suggest that both type of 
quantum criticality will strongly influence the development of the dome of high temperature superconductivity in its vicinity.

{\it Acknowledgement.---} 
The work was supported in part by the 
U.S. Department of Energy, Office of Science,
Basic Energy Sciences, under Award No.\ DE-SC0018197
 (W.-J.H., H.-H.L., and Q.S.), 
the Robert A.\ Welch Foundation Grant No.\ C-1411 (H.H., and Q.S.), 
a Smalley Postdoctoral Fellowship of the Rice Center for Quantum Materials (H.-H. L.),
the National Science Foundation of China Grant number 11674392
and Ministry of Science and Technology of China,
National Program on Key Research Project Grant number 2016YFA0300504 (R.Y.).
The majority of the computational calculations have been 
performed on the Shared University Grid at Rice funded by NSF under Grant EIA-0216467, 
a partnership between Rice University, Sun Microsystems, and Sigma Solutions, 
Inc., the Big-Data Private-Cloud Research Cyberinfrastructure MRI-award funded by NSF under Grant No. CNS-1338099 
and by Rice University, the Extreme Science and 
Engineering Discovery Environment (XSEDE) by NSF under Grant No.\ DMR160057. 
Q.S. acknowledges  the hospitality and support by a Ulam Scholarship
of the Center for Nonlinear Studies at Los Alamos National Laboratory
and
the hospitality of the Aspen Center for Physics (NSF grant No. PHY-1607611).

\bibliography{multiorbital_iron}

\clearpage

\begin{center}
\begin{large}
{\bf Supplemental Material}
\end{large}
\end{center}

\section{The filling number of each orbital for iron pnictides}
\begin{figure}[h!]
\includegraphics[width=0.60\columnwidth]{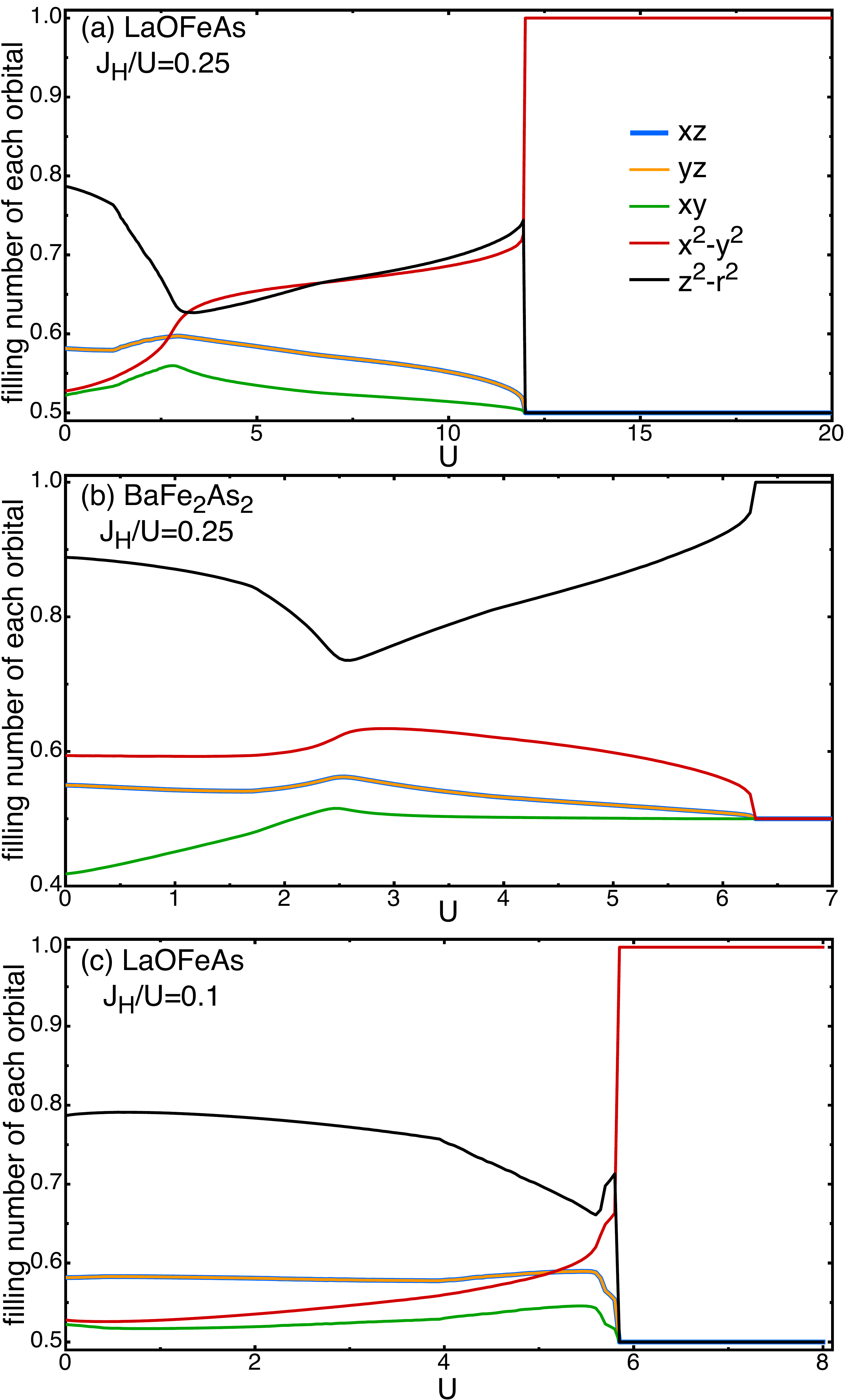}
\caption{The slave-spin results of filling number in the 5-orbital Hubbard model for LaOFeAs with the Ising-only
Hund's coupling $J_H=0.25U$ (a) and for BaFe$_{2}$As$_2$ and Ising-only $J_H=0.25U$ (b), as well as for LaOFeAs with the 
Ising-only Hund's coupling $J_H=0.1U$ (c).}
\label{band}
\end{figure}

To estimate the electron filling in each orbital, we have performed calculations by using the U(1) slave-spin method~\cite{Yu.12a}
on 5-orbital Hubbard models for iron pnictides LaOFeAs and BaFe$_{2}$As$_2$ with six electrons occupying the $5$ $3d$-orbitals of each Fe ion.
The results for the Ising-only Hund's coupling $J_H/U=0.25$ for LaOFeAs and BaFe$_{2}$As$_2$ are shown in Fig.\,\ref{band}(a,b).
For large interactions in each model, one of the two $e_g$ orbitals is doubly occupied ($x^2-y^2$ orbital for LaOFeAs and $z^2-r^2$ 
orbital for BaFe$_{2}$As$_2$), while all the other orbitals are at half filling. Most importantly, for both models, at small and intermediate interactions, 
the filling numbers of the
three $t_{2g}$ orbitals $xz$, $yz$, and $xy$ are close to half filling. These results are consistent with those in our VMC calculations
shown in Fig.\,4 and justifies setting the total filling number to half-filling in the 3-orbital Hubbard model in the VMC calculations.
We have used a larger Ising-only Hund's coupling to mimic the effect of SU(2)-invariant Hund's coupling of $J_H/U=0.1$, but our conclusion
is not sensitive to the precise value of $J_H/U$. We illustrate this point in Fig.\,\ref{band}(c), with a choice of the Ising-only $J_H/U=0.1$;
here again, 
the filling numbers of the
$xz$, $yz$, and $xy$ orbitals are close to half filling for small and intermediate interactions.

\section{Additional numerical results by variational Monte Carlo}
\begin{figure}
\includegraphics[width=\columnwidth]{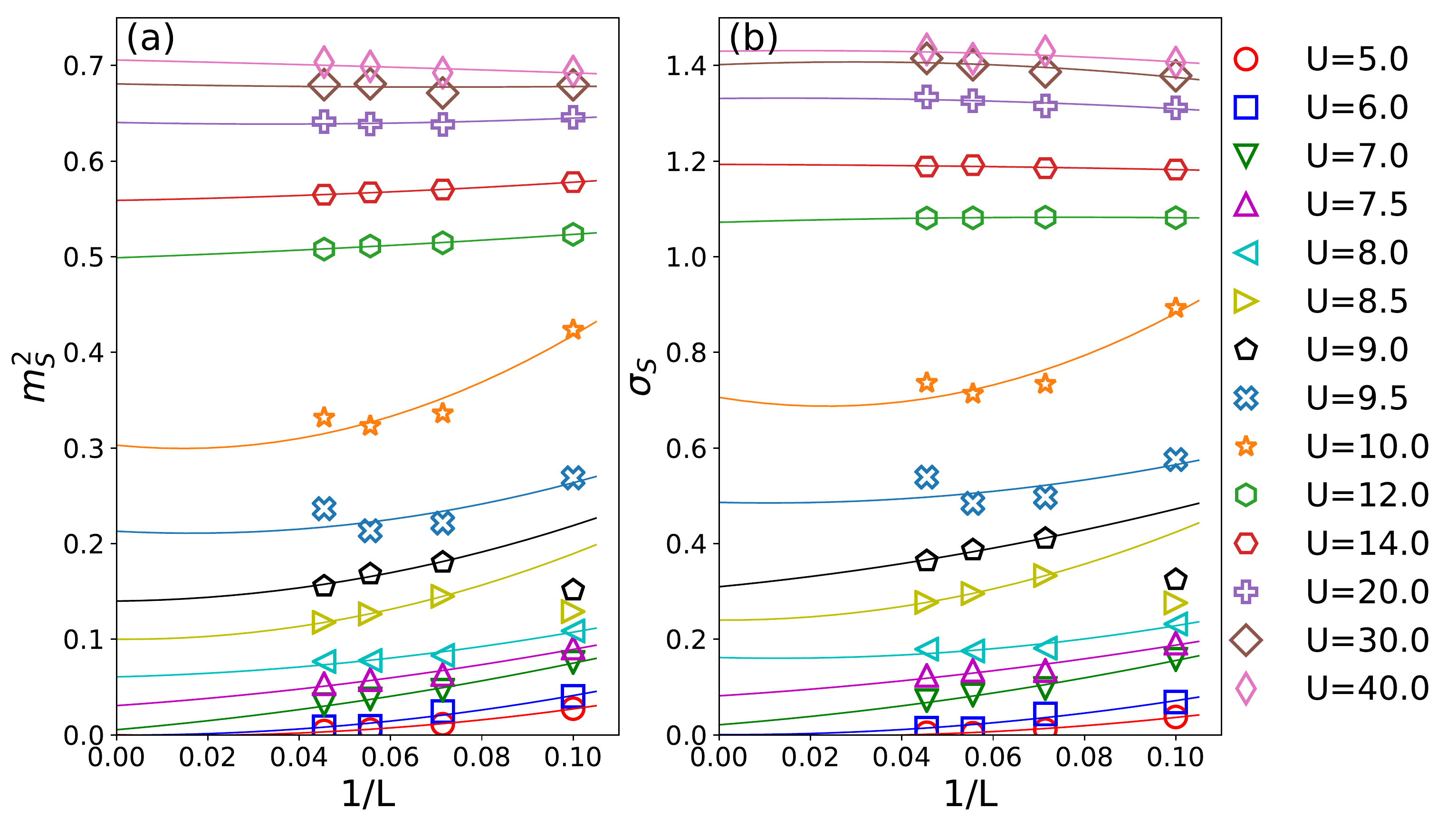}
\includegraphics[width=\columnwidth]{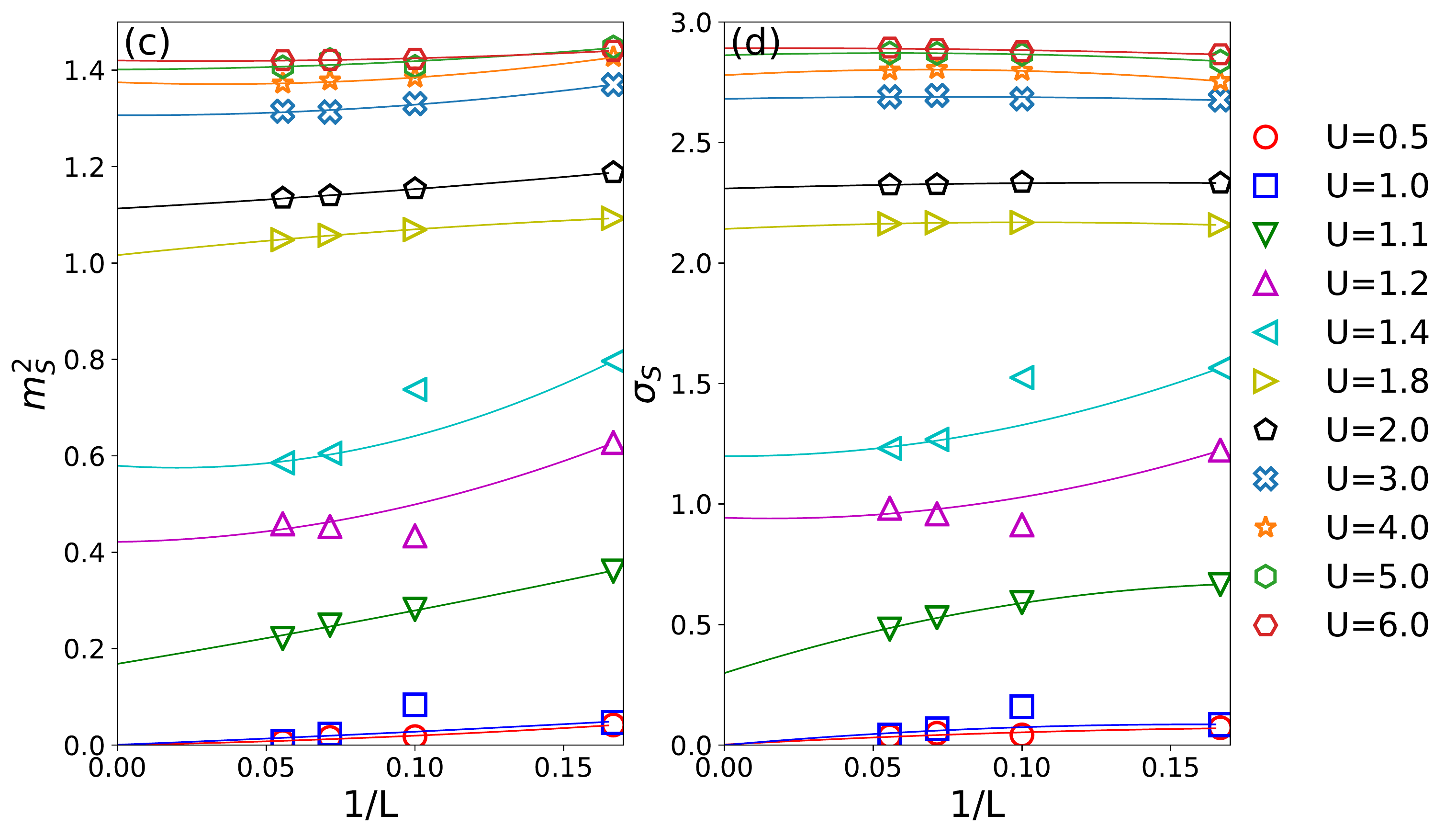}
\caption{(a) and (b) The finite size scaling for the magnetic $m^2_S$ and nematic $\sigma_S$ order parameters at different $U$ for the 
2-orbital Hubbard model. (c) and (d) The finite size scaling for the magnetic $m^2_S$ and nematic $\sigma_S$ order parameters at different
$U$ for the 3-orbital Hubbard model. A quadratic fitting has been performed, with the lines being the best fitting results.}
\label{fit}
\end{figure}

In Fig.~\ref{fit}, we present the finite size scaling for the magnetic $m^2_S$ and the nematic $\sigma_S$ order parameters at different 
interactions $U$ in both the 2- and 3-orbital Hubbard model with the Hund's coupling $J_H/U=0.1$. For the 2-orbital Hubbard model, 
at small interactions, i.e., $U<7.5$, the finite size scaling show vanishing values of $m^2_S$ and $\sigma_S$ in the thermodynamic limit; 
at $U\simeq 7.5$, small but nonzero
 $m^2_S$ and $\sigma_S$ suggest concurrent magnetic and nematic phase transitions. For the 3-orbital 
Hubbard model, nonzero $m^2_S$ and $\sigma_S$ after the finite size scaling are obtained around $U=1.0$, suggesting again concurrent 
magnetic and nematic phase transitions. We notice that the extrapolation to the thermodynamic limit is remarkably smooth, except in the 
metallic phase with CAFM order, where stronger fluctuations of the order parameters with the system size are observed. Nevertheless, a 
finite value of the order parameters in the thermodynamic limit, monotonically increasing with the interaction $U$, can be clearly 
obtained also in the metallic phase with CAFM order.

\begin{figure}
\includegraphics[width=\columnwidth]{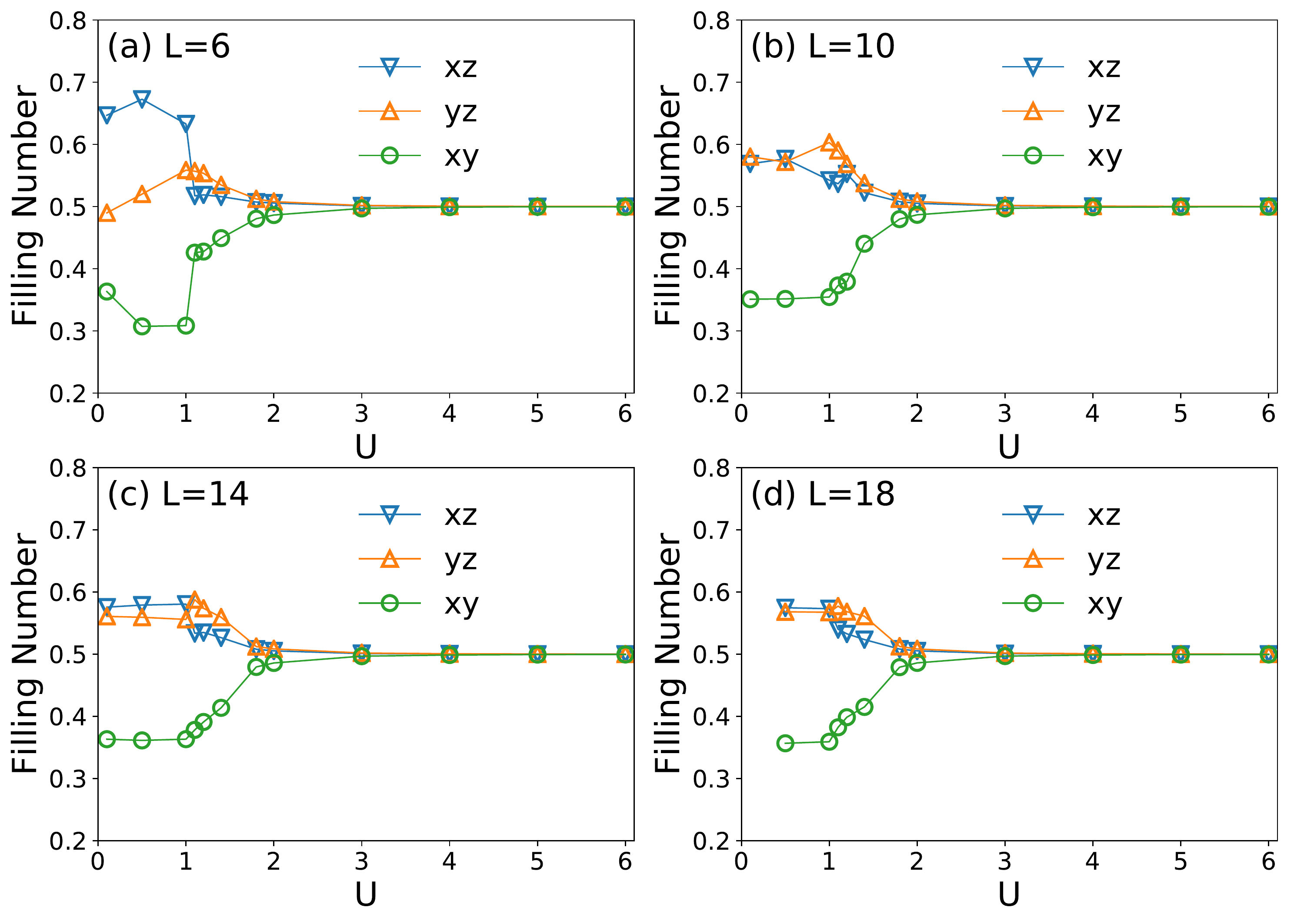}
\caption{The filling numbers of $xz$, $yz$, and $xy$ orbitals on $L=6$ (a), $10$ (b), $14$ (c), and $18$ (d) clusters for the the 
3-orbital Hubbard model.}
\label{fillingeach}
\end{figure}

For the 3-orbital Hubbard model, we have calculated additional data for the filling numbers of each orbital, see Fig.~\ref{fillingeach}.
It clearly shows the difference between the behavior of the orbital $xy$ from orbitals $xz$ and $yz$, especially at small and intermediate
interactions.

\end{document}